\documentclass[10pt,letterpaper]{article}
\usepackage{opex3}
\usepackage{hyperref}

\begin{document}
\title{Single-laser, one beam, tetrahedral magneto-optical trap}

\author{Matthieu Vangeleyn, Paul F. Griffin, 
 Erling Riis, Aidan S. Arnold}

\address{Department of Physics, SUPA, University of Strathclyde, Glasgow G4 0NG, UK}

\homepage{http://www.photonics.phys.strath.ac.uk/AtomOptics/} 


\begin{abstract}
We have realised a 4-beam pyramidal magneto-optical trap ideally suited for future microfabrication. Three mirrors split and steer a single incoming beam
into a tripod of reflected beams, allowing trapping in the four-beam overlap volume. We discuss the influence of mirror angle on cooling and trapping,
finding optimum efficiency in a tetrahedral configuration. We demonstrate the technique using an ex-vacuo mirror system to illustrate the previously
inaccessible supra-plane pyramid MOT configuration. Unlike standard pyramidal MOTs both the pyramid apex and its mirror angle are non-critical and our
MOT offers improved molasses free from atomic shadows in the laser beams. The MOT scheme naturally extends to a 2-beam refractive version with high
optical access. For quantum gas experiments, the mirror system could also be used for a stable 3D tetrahedral optical lattice.
\end{abstract}



\section{Introduction}
The last decade has seen rapid progress in the development of microfabricated atom `chip' traps. With these devices one can expect to achieve small atom
number trapping, opening new perspectives towards e.g.\ cavity quantum electrodynamics (QED) experiments \cite{Stamper-Kurn:2008:a} and ultimately the
realization of the quantum computer \cite{Walther:2007:a}. In surface atom chips, the very strong magnetic field gradients produced by the micron-sized
conductive wires \cite{Schmiedmayer:1999:a} or by permanent magnets \cite{Hinds:2005:a} provides very tight confinement for accurate manipulation of
magnetically trapped atoms. Bose-Einstein condensates (BECs) can be very quickly obtained given the high trapped frequencies leading to a dramatic
increase of the RF evaporative cooling stage efficiency, and therefore experiment time can be greatly reduced.

Another very attractive way to trap atoms on a chip is achieved using micromirrors \cite{edpyra} etched in a 4-sided pyramidal shape \cite{Lee:1996:a}
and cooling atoms in a magneto-optical trap (MOT) - a design experimentally realised only very recently \cite{edprep}. The pyramid constitutes an ideal
technique for generating e.g.\ slow atomic beams but the geometry of the device dictates that the MOT is situated within the volume of the pyramid.
Hence, any further manipulation of the cold atoms is strongly limited by the absence of any direct optical access.

\begin{figure}[!t] \centering
\begin{minipage}{0.48\columnwidth}
\centering\includegraphics[width=.99\columnwidth]{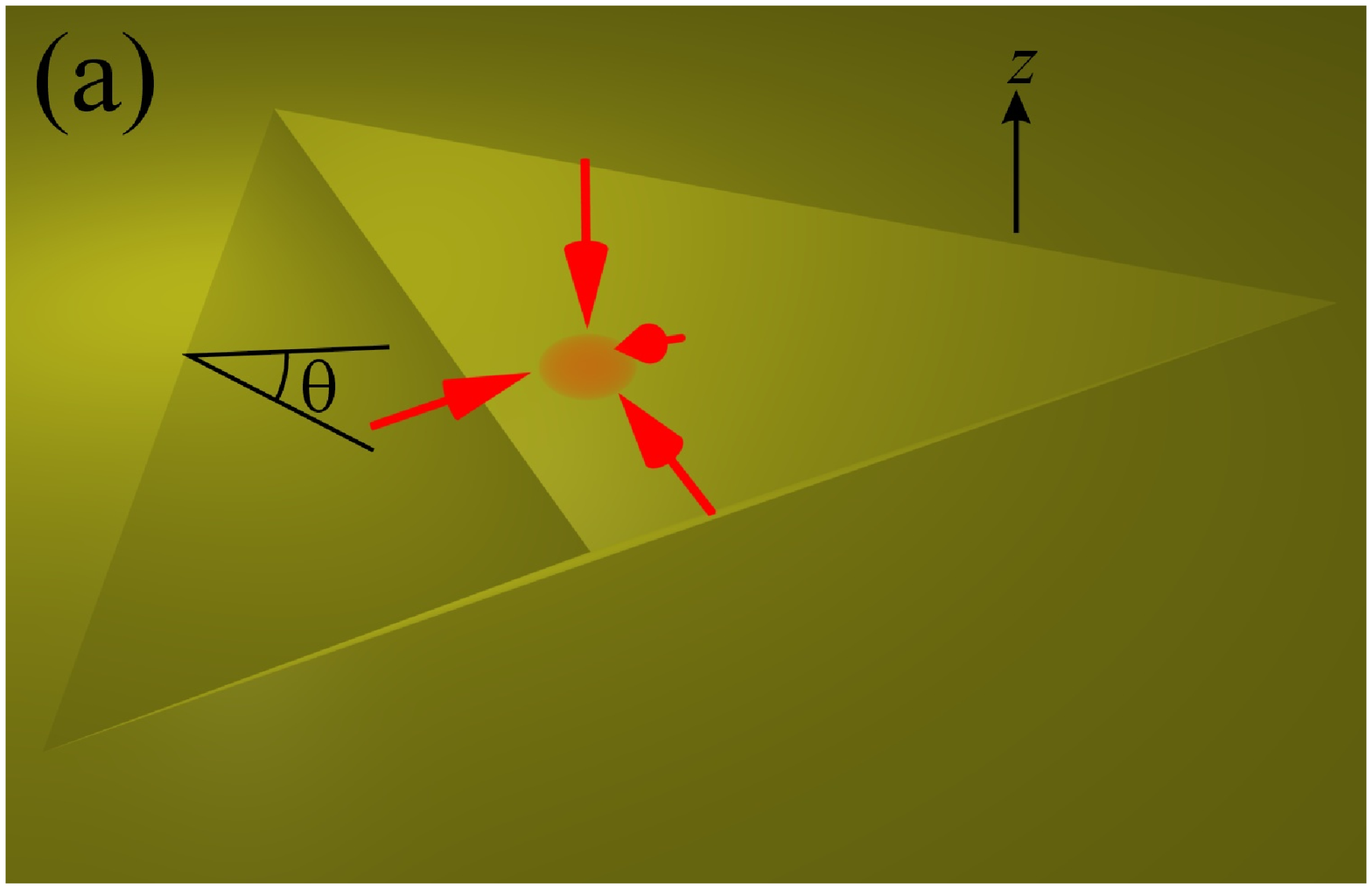}
\end{minipage}
\begin{minipage}{0.45\columnwidth}
\centering\includegraphics[width=.99\columnwidth]{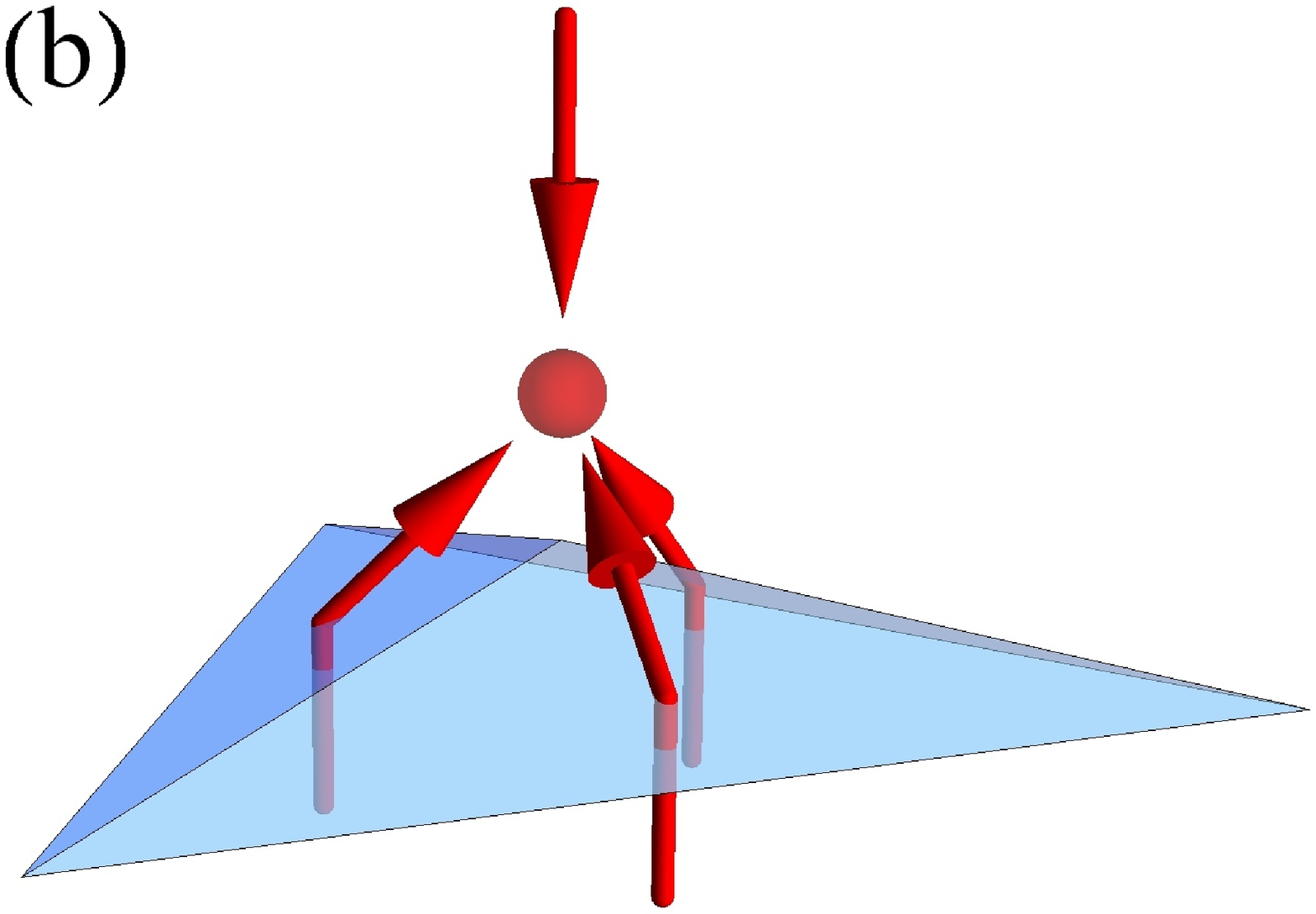}
\end{minipage}
\caption{\label{fig1} Schematic of the ideal reflective version of the tetrahedral pyramid MOT (a). A single downwards laser beam is split and reflected
by a pyramid of mirrors. Mirror declination of $\theta$ relative to the horizontal plane yields reflected beams at angle $\pi/2-2\theta$ above the plane.
A refractive version (b) would have independent upward and downward laser beams, here with upward beams at angle $\pi/4$ above the plane, using a diamond
$(n=2.4)$ pyramid with inclination $\theta=22.7^{\circ}$. In both cases balanced optical molasses is formed when the intensity-weighted $\textbf{k}$
vectors of the four beams add to zero, i.e. $I_{\rm up}=I_{\rm down}/(3 \cos 2\theta$).}
\end{figure}

A minimum of two focused beams can form a MOT, however cooling and trapping forces are relatively weak \cite{Tabosa}. Shimizu's group experimentally
showed that one can obtain a robust 3D MOT using four collimated laser beams \cite{Shimizu:1991:a}.  We propose and demonstrate a novel pyramidal version
of the tetrahedral four beam MOT using only a single incident beam reflected by a set of three mirrors (Fig.~\ref{fig1} (a)).
   The tetrahedron is
readily scalable to smaller dimensions for a microfabricated (e.g.\ focused ion beam) atom trap and has the additional advantage that the beam overlap
region extends above the surface, allowing supra-plane operation. Atoms, while still in a MOT cooling stage, can therefore be easily shifted (using a
constant magnetic field to manipulate the trap center) and addressed.

One can consider a standard pyramidal MOT as `9' separate beams: an incident beam is split by the four triangular pyramid mirrors into 4 horizontal
triangular beams, which are subsequently reflected into 4 triangular beams (ideally) counterpropagating with the incident beam. Incorrect pyramid apex
angle will result in a bright or dark `cross' in the trap, and beam intensity variations will be exacerbated by the sharp-edge diffraction on the sides
and particularly the apex of the pyramid. The apex of our pyramid is not required for MOT operation and it can be physically removed for e.g.\ absorption
imaging, or using the MOT as a cold atom beam source. Moreover, diffraction from the sides of our pyramid will only affect the edges of the capture
volume, far from the MOT.

We first clarify the theoretical properties of the MOT in section~\ref{theorysec}, extending previous descriptions, then detail our experimental
observations in section~\ref{exptsec}. In section~\ref{speculate} we highlight a refractive pyramid geometry (Fig.~\ref{fig1} (b)), and note that a
tetrahedral beam intersection (via either a reflective or refractive 3-sided pyramid) also provides an ideal tool for a high phase stability optical
lattice, with the benefit of fixed lattice geometry.

\section{Theory}\label{theorysec}

The acceleration, due to light scattering from laser beam `$j$' propagating in the direction of unit vector $\hat{\textbf{k}}_j$, on a single atom with
velocity $\textbf{v}$ in a magnetic field $\textbf{B}$ is (e.g.\ \cite{lindquist,foot}):
\begin{equation}
 \textbf{a}_j=a \, \beta_j \, \hat{\textbf{k}}_j \sum_{n=-1,0,1}
    \eta_{n}/(1+\beta_{\rm tot}+4\,(\Delta_\Gamma-k_\Gamma \, \hat{\textbf{k}}_j\cdot\textbf{v}-\mu_\Gamma \, n \, |\textbf{B}|)^2),
\label{acc}\end{equation} where $a= h \, \pi \, \Gamma/ (m \, \lambda),$ beam $j$ has intensity $\beta_j=I_j/I_S$ relative to the saturation intensity
and $\beta_{\rm tot}=\sum_j{\beta_j}$. Note $\eta_n$ describes the relative light polarisation, detailed below. For the specific case of $^{87}$Rb atoms,
the atomic linewidth is $\Gamma=6.07\,$MHz, $\lambda=780\,$nm, $a=111\,\textrm{km}\,\textrm{s}^{-2}$, and $I_S=1.67\,\textrm{mW}/\textrm{cm}^{2}$. The
detuning of the laser is $\Delta_\Gamma$ in units of $\Gamma$, $k_\Gamma=\lambda^{-1}\Gamma^{-1}$ and $\mu_\Gamma=\mu_B/(2\,\pi\,\Gamma).$ Note we have
assumed an $F=0$ to $F=1$ atomic transition purely to simplify the mathematics. For  $k_{\Gamma}|\textbf{v}|,\,\mu_\Gamma|\textbf{B}|\ll \Delta_\Gamma$,
we can Taylor expand the denominator in Eq.~\ref{acc} to obtain
\begin{equation}
 \textbf{a}_j \approx a \, \beta_j \, \hat{\textbf{k}}_j \sum_{n=-1,0,1}
    {\eta_{n}\, (K + C\, (k_\Gamma \, \hat{\textbf{k}}_j\cdot\textbf{v}+\mu_\Gamma \,n \,|\textbf{B}|))}\, ,
\label{acc2}\end{equation} where $K=(1+\beta_{\rm tot}+4\,{\Delta_\Gamma}^2)^{-1}$ and $C=8\,\Delta_\Gamma \, K^2$.

The position dependence of the acceleration is characterized by the projection of the beam propagation direction onto the local magnetic field, i.e.
$\zeta=\hat{\textbf{k}}_j\cdot\hat{\textbf{B}}$, which determines the decomposition of circularly polarised light with handedness $s=\pm 1$ (relative to
the propagation direction) into its relative $\{\sigma_{-},\pi,\sigma_{+}\}$ polarised components $\eta_n$ (where $n=-1,0,+1$ respectively). One can show
that $\eta_{0}=(1-\zeta^2)/2$ and $\eta_{\pm1}=(1\mp s \, \zeta)^2/4,$ thus $\sum_n{\eta_n}=1$, and we can expand Eq.~\ref{acc2} to obtain:
\begin{equation}
 \textbf{a}_j\approx a\, \beta_j \,\hat{\textbf{k}}_j \,(K + C\, (k_\Gamma \, \hat{\textbf{k}}_j \cdot \textbf{v}-s \,\zeta \,\mu_\Gamma \,|\textbf{B}|))\, .
\label{acc3}\end{equation} If we assume a spherical quadrupole magnetic field with symmetry axis along $z$, i.e. $\textbf{B}=b \{x,y,-2z\},$ then a beam
with $s=-1$ and $\hat{\textbf{k}}=\{1,0,0\}$ will thus generate an acceleration $\textbf{a}=a \, \beta_j \, \hat{\textbf{k}}_j (K+C \, (k_\Gamma \, v_x
+\mu_\Gamma \, b \, x)).$ To obtain a balanced magneto-optical trap and optical molasses we require a zero net acceleration $\textbf{a}\propto
\sum_{j}{\beta_j \hat{\textbf{k}}_j}=\textbf{0}.$ The standard 6-beam MOT is obtained using $\beta_j=\beta=\beta_{\rm tot}/6$ for all beams, $s=1$ for
the $\pm z$ beams and $s=-1$ for the $\pm x$ and $\pm y$ beams, yielding a cylindrically symmetric acceleration $\textbf{a}_{\rm
tot}=\sum_{j}{\textbf{a}_j}=2 a \, \beta \, C ( k_\Gamma\, \{v_r,v_z\}+\mu_\Gamma \,b \,\{r,2z\}),$ (which traps and cools with red-detuned
$(\Delta_\Gamma<0)$ light as $C$ is proportional to $\Delta_\Gamma$).

\begin{figure}[!b]
\centering\includegraphics[width=.5\columnwidth]{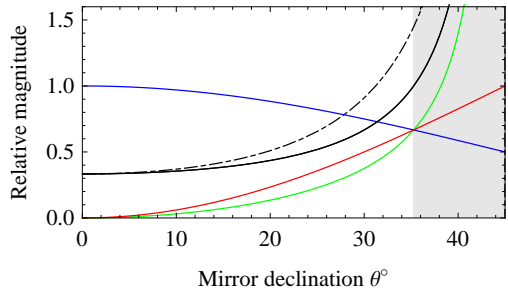}
 \caption{\label{figth} Axial cooling (blue), axial trapping (red) and radial
trapping/cooling (green) forces, as a function of the mirror declination, $\theta$ (Eq.~\ref{acc4}). Forces are shown relative to radial forces in the
6-beam MOT. Ideal relative reflected beam power $I_{\rm up}/I_{\rm down}=1/(3\cos 2\theta)$ is shown in black, and the maximum reflected beam power
(dash-dotted curve) is discussed in the text. Balanced molasses in the grey zone at angles above the pure tetrad $(\theta=35.3^\circ)$ can only be
obtained in the refractive geometry with independent control of $I_{\rm up}$ and $I_{\rm down}$.}
\end{figure}

The usual MOT uses six orthogonal beams because vacuum cells that are cuboidal or have perpendicular viewports are easily manufactured, and alignment is
easier when retroreflections are used. However, in order to get a point where there is still strong 3D cooling and the total radiation pressure cancels,
the minimum number of laser beams is four. If these four beams have equal intensity, then the beams cross in a tetrahedral configuration, in which each
pair forms an angle of $\arccos{(-1/3)}\approx 109.5^\circ$ (i.e. the mirror declination is $\theta=35.3^\circ$). In our case, only one beam (in the $-z$
direction with $s=1$) is brought to the cell with intensity $I_{\rm down}$ and three mirrors are used to split and reflect the incoming beam into three
upward beams with intensity $I_{\rm up}$ and $s=-1$. The upward beam circular handedness changes $s=1\rightarrow s=-1$ on reflection from the mirrors of
the pyramid. The upward beams require an intensity $I_{\rm up}=I_{\rm down}/(3\cos 2\theta)$ for balanced optical molasses and hence efficient
sub-Doppler cooling (this corresponds to an intensity loss on reflection for mirror angles smaller than the pure tetrad). The total acceleration:
\begin{equation}\label{acc4}
\textbf{a}_{\rm tot}=\sum_{j}{\textbf{a}_j}=(a \, \beta C/2) (k_\Gamma \, \{\sin 2\theta \tan 2\theta v_r,4\cos^2\theta v_z\}+ \mu_\Gamma b \{\sin
2\theta \tan 2\theta r,8 \sin^2\theta z\}), \end{equation} is depicted in Fig.~\ref{figth} which shows the relative cooling and restoring acceleration at
the MOT formation point as a function of mirror declination.

\begin{figure}[!t]
\begin{minipage}{0.49\columnwidth}
\centering\includegraphics[width=.90\columnwidth]{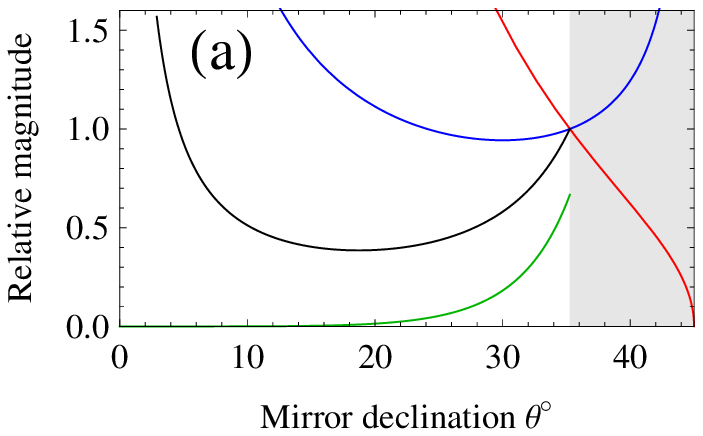}
\end{minipage}
\begin{minipage}{0.49\columnwidth}
\centering\includegraphics[width=.99\columnwidth]{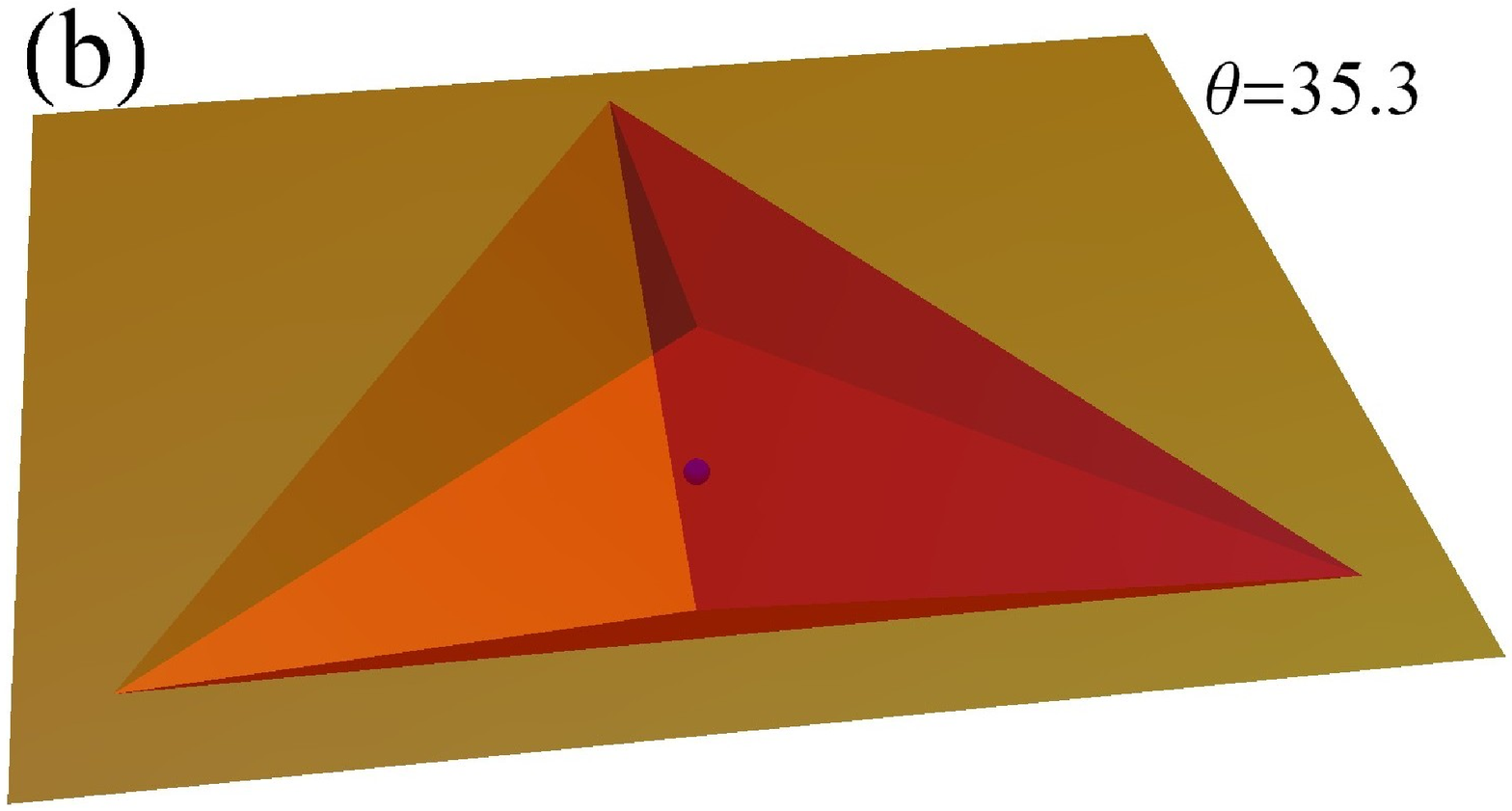}
\end{minipage}
\caption{\label{volfig}Image (a) show the total (black) and sub-plane (green) MOT capture volume, respectively, relative to the pure tetrad capture
volume. Also shown are the relative axial (blue) and radial (red) MOT sizes. The beam intersection for $\theta=35.3^{\circ}$ is illustrated in (b), and
the $\theta$-dependent capture volume (and relative MOT \textit{size} (blue ellipsoid)) can be observed as an animation
(\href{http://www.photonics.phys.strath.ac.uk/AtomOptics/Tetra.html}{www.photonics.phys.strath.ac.uk/AtomOptics/Tetra.html}).}
\end{figure}

MOT operation is possible for almost any wedge angle however confinement and radial cooling vanish as $\theta \rightarrow 0.$ The low-intensity relative
size of the final MOT in the $i$ direction (Fig.~\ref{volfig}) is given by $\sigma_{i}\propto\sqrt{I_{\rm tot}}/\sqrt{\gamma_i \kappa_i}$ where $I_{\rm
tot}=3 I_{\rm up}+I_{\rm down}\propto 1+\sec2\theta$ and $\gamma_i,$ $\kappa_i$ are the relative damping and restoring constants from Eq.~\ref{acc4}.
Note that small MOT \textit{size} is desirable, as the small volume of the MOT itself is a result of optimal cooling and trapping forces, implying a
large atom number capture and hence a high density atom cloud.

The overlap volume of the four beams determines the number of atoms in the MOT, and is the intersection of two triangular pyramids (the lower pyramid
fills the mirror pyramid) in the case of a pure tetrad (Fig.~\ref{volfig}(b)). More generally the overlap volume is the intersection of a hexagonal
pyramid and a triangular pyramid, this is depicted in the movie link from Fig.~\ref{volfig}. With appropriate manipulation one finds the total and
sub-plane capture volumes have the simple trigonometric relation $V_{\rm tot}\propto 1/(\sin4\theta (1+3\cos 2\theta)),$ $V_{\rm in}\propto 3\sin^2\theta
\tan^3\theta/(2\cos2\theta).$ By applying an appropriate constant magnetic field vector in addition to the magnetic quadrupole it is possible to form a
MOT anywhere within the beam overlap volume. The overlap volume tends to infinity for small $\theta,$ however the reduced radial cooling precludes use as
an extended `beam MOT.' For angles $\theta>35.3^\circ$ parts of the upward beams reflect twice from the pyramid, which reduces the capture volume of the
MOT and balanced optical molasses is impossible (at least for the reflection pyramid MOT). These double-reflected ($s=1$) beams yield radially
anti-trapping and axially trapping forces.

\begin{figure}[!h]
\begin{minipage}{0.302\columnwidth}
\centering\includegraphics[width=.99\columnwidth]{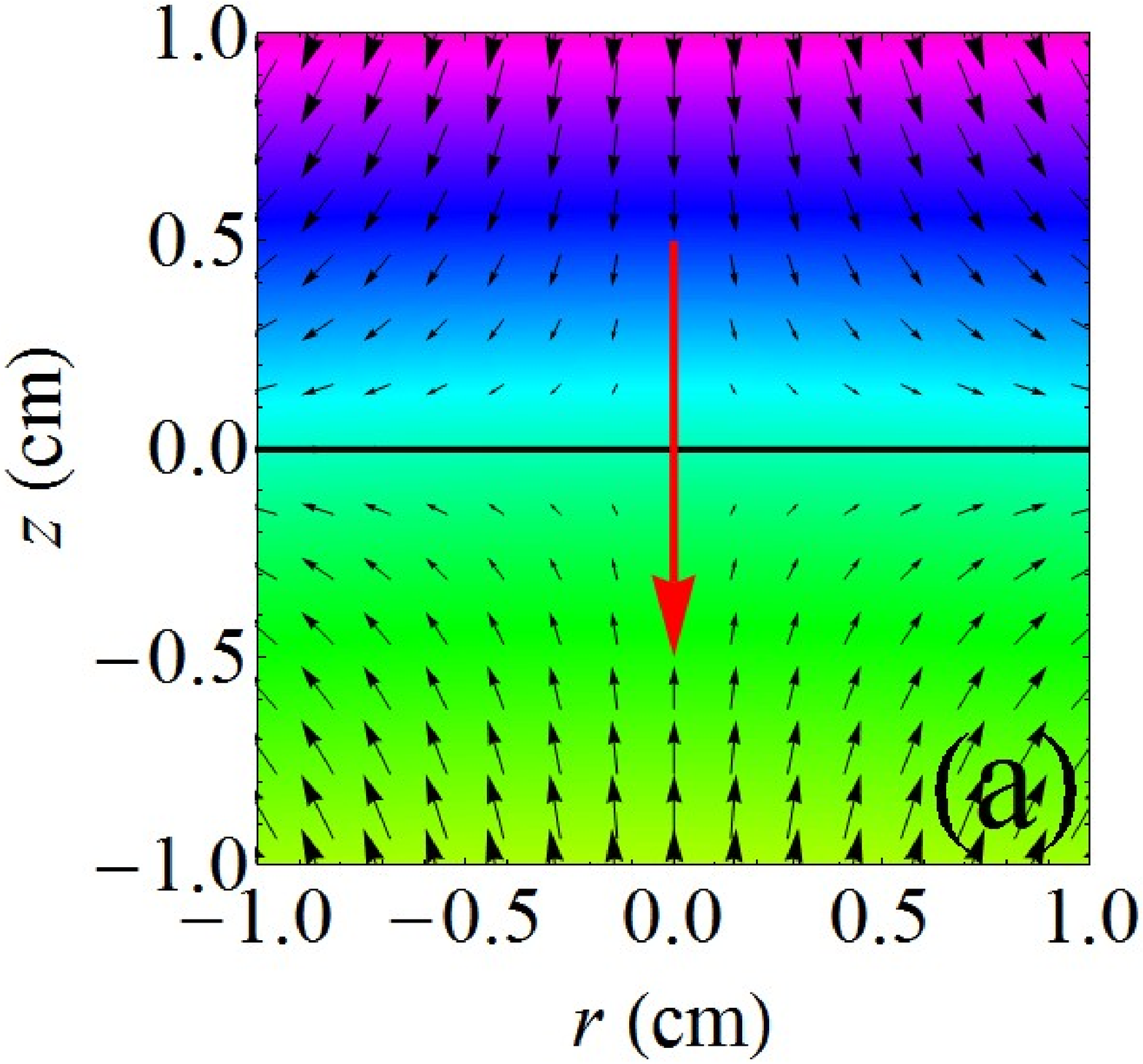}
\end{minipage}
\begin{minipage}{0.302\columnwidth}
\centering\includegraphics[width=.99\columnwidth]{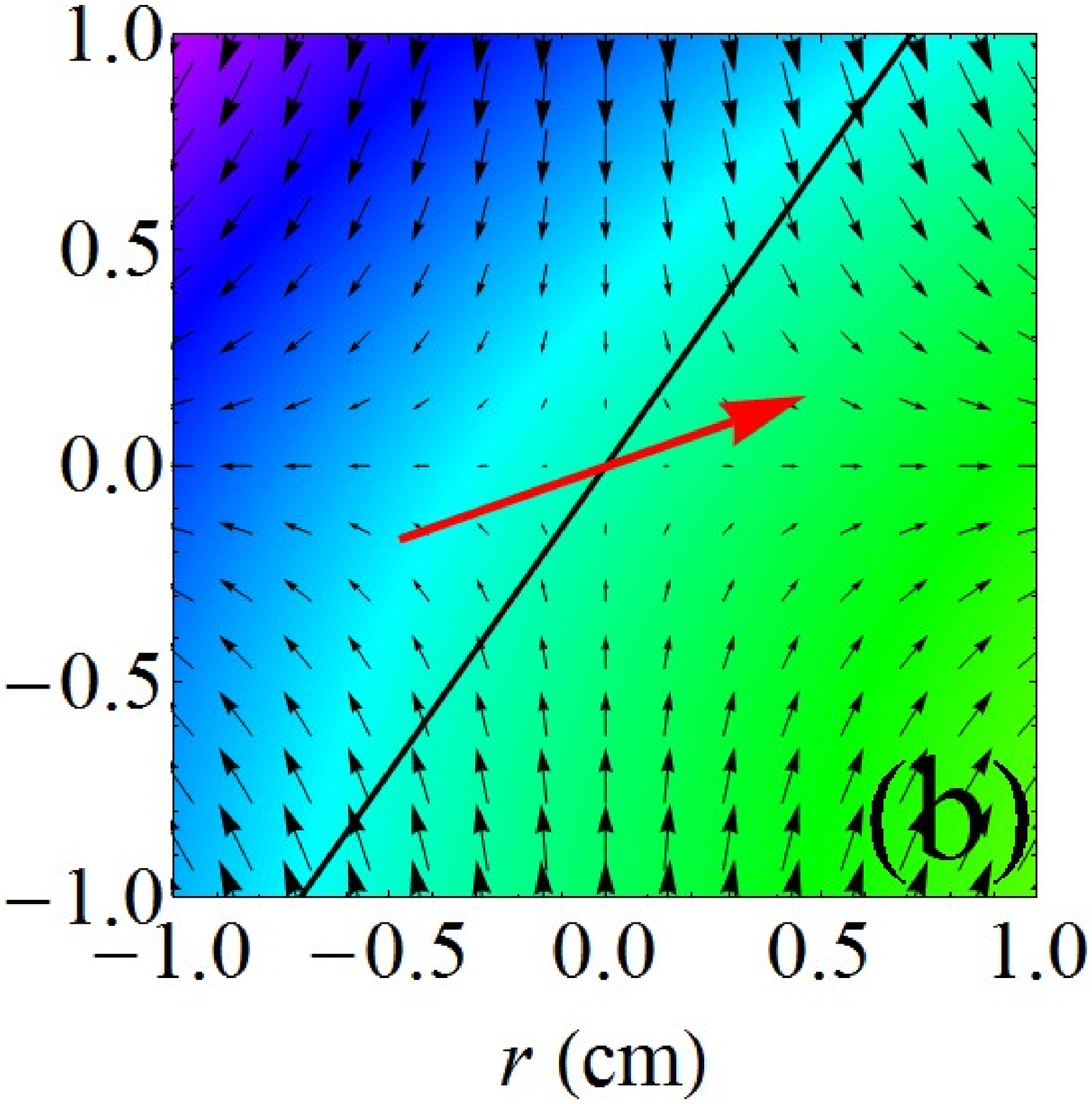}
\end{minipage}
\begin{minipage}{0.302\columnwidth}
\centering\includegraphics[width=.99\columnwidth]{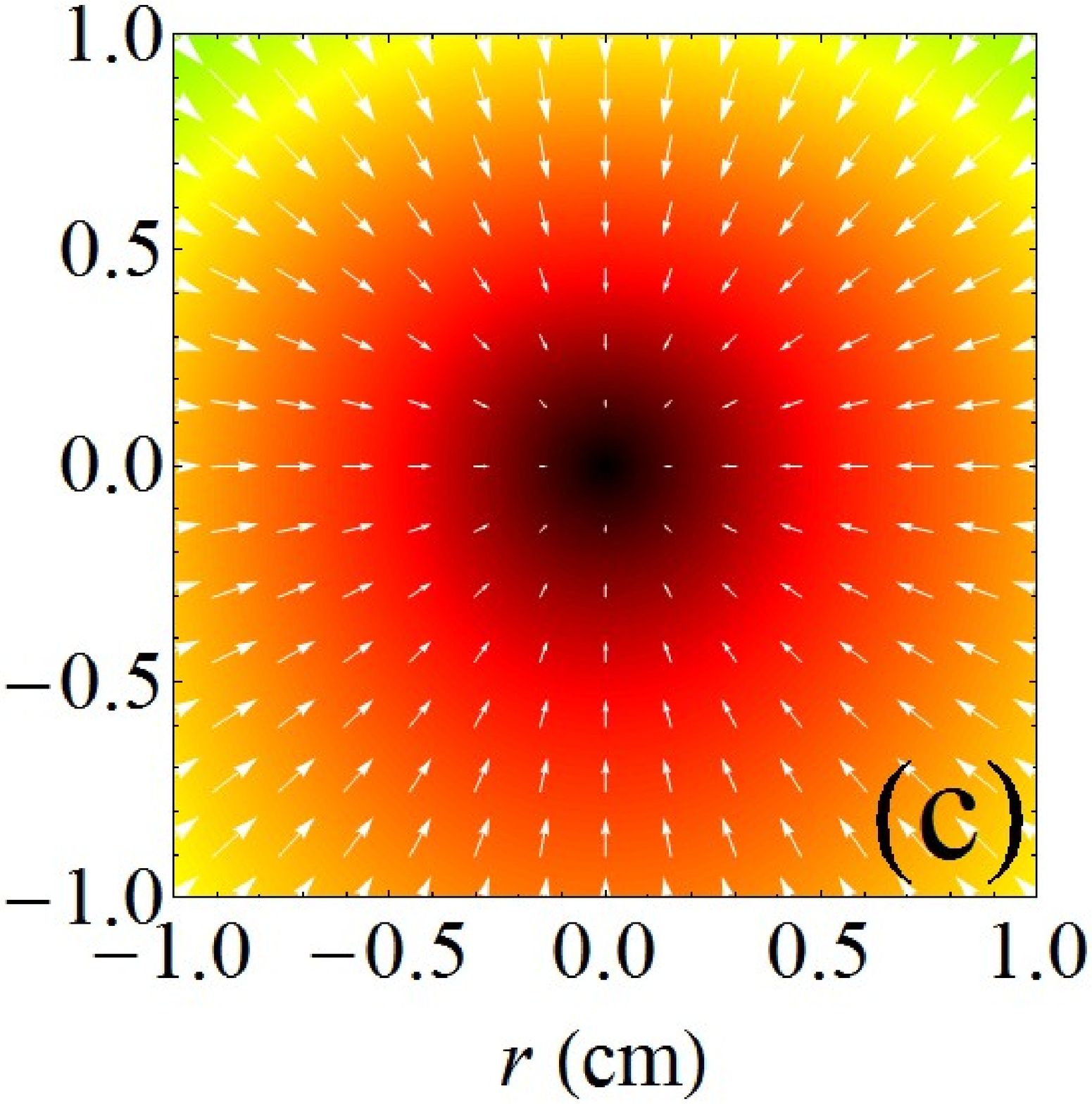}
\end{minipage}
\begin{minipage}{0.075\columnwidth}
\centering\includegraphics[width=.99\columnwidth]{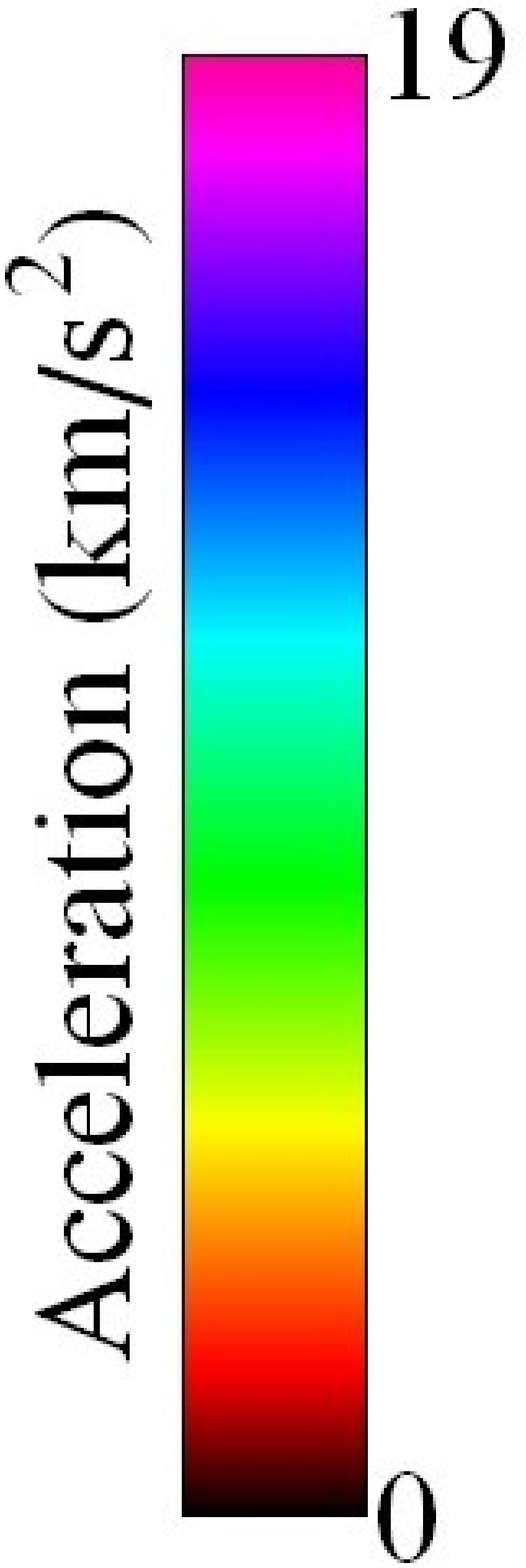}
\end{minipage}
 \caption{\label{accfig} Images (a) and (b) are the acceleration magnitudes
due to single laser beams in the direction of the red arrows -- (b) corresponds to radial trapping and weak axial \textit{anti}-trapping. The circular
polarisation changes from $s=1$ in (a) to $s=-1$ in (b) due to the reflection from the pyramid mirror.  The local magnetic field and the
$\textbf{k}\cdot\textbf{B}=0$ line are shown as black vectors and a black line, respectively. The acceleration due to a single upward beam (b) as
$\theta$ varies can be seen at:
\href{http://www.photonics.phys.strath.ac.uk/AtomOptics/Tetra.html}{www.photonics.phys.strath.ac.uk/AtomOptics/Tetra.html}. Image (c) shows the
acceleration magnitude (colorscale) and vectors (white) due to all four tetrahedral MOT beams. Acceleration is isotropic, whereas standard MOTs have
$a_r:a_z=1:2$.}
\end{figure}
The full acceleration in the pyramidal MOT (Eq.~\ref{acc}), broken down into contributions from the `up' and `down' beams is shown in Fig.~\ref{accfig}.
We used parameters similar to those in our experimental demonstration in the next section, i.e.\ $\beta_{\rm down}=1.38 (=\beta_{\rm up}),$
$\Delta_\Gamma=-1.43,$ and a magnetic field gradient of $b=10\,$G/cm. The images in Fig.~\ref{accfig} use balanced intensities, however one critical
feature of the tetra-MOT is that although the upwards beams are radially trapping, they are increasingly axially anti-trapping as $\theta\rightarrow 0$.
If the upwards beams are sufficiently intense, the anti-trapping effect cancels the downward beam's axial trapping, and it is impossible for a MOT to
form. This critical intensity happens when $a_{\rm tot}(z_0)=0$ and $a_{\rm tot}'(z_0)=0$ (a saddle-node bifurcation), which we simultaneously solve for
$I_{\rm up}/I_{\rm down}$ and the position $z_0$. MOTs cannot form above a critical ratio of $I_{\rm up}/I_{\rm down}$ (the black dash-dotted line in
Fig.~\ref{figth}), and there will be further restrictions based on whether $z_0$ is within the MOT beam overlap.

\section{Experiment}
\label{exptsec}

\begin{figure}[!b]
\centering
\begin{minipage}{0.45\columnwidth}
\centering\includegraphics[width=.99\columnwidth]{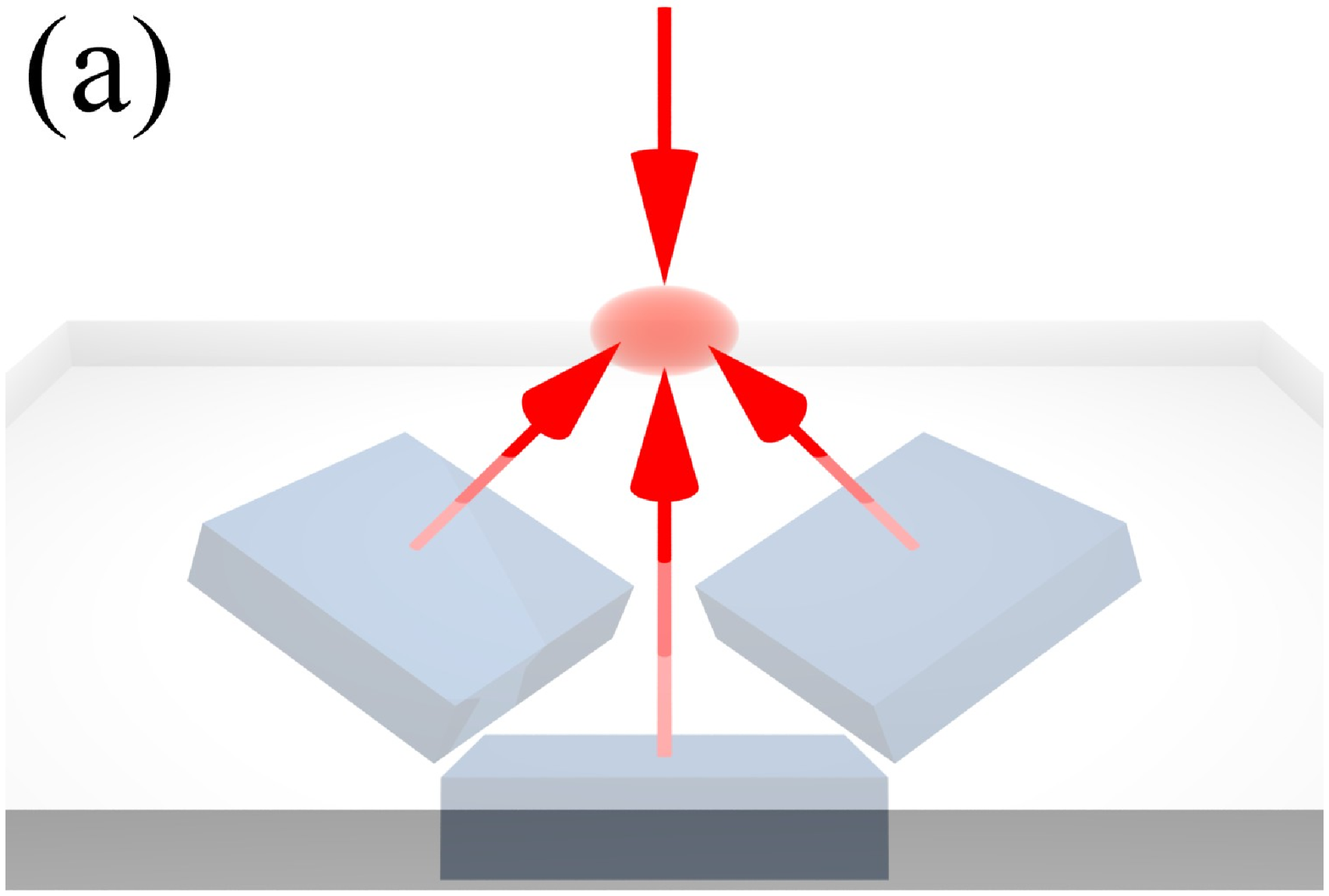}
\end{minipage}
\begin{minipage}{0.4\columnwidth}
\centering\includegraphics[width=.99\columnwidth]{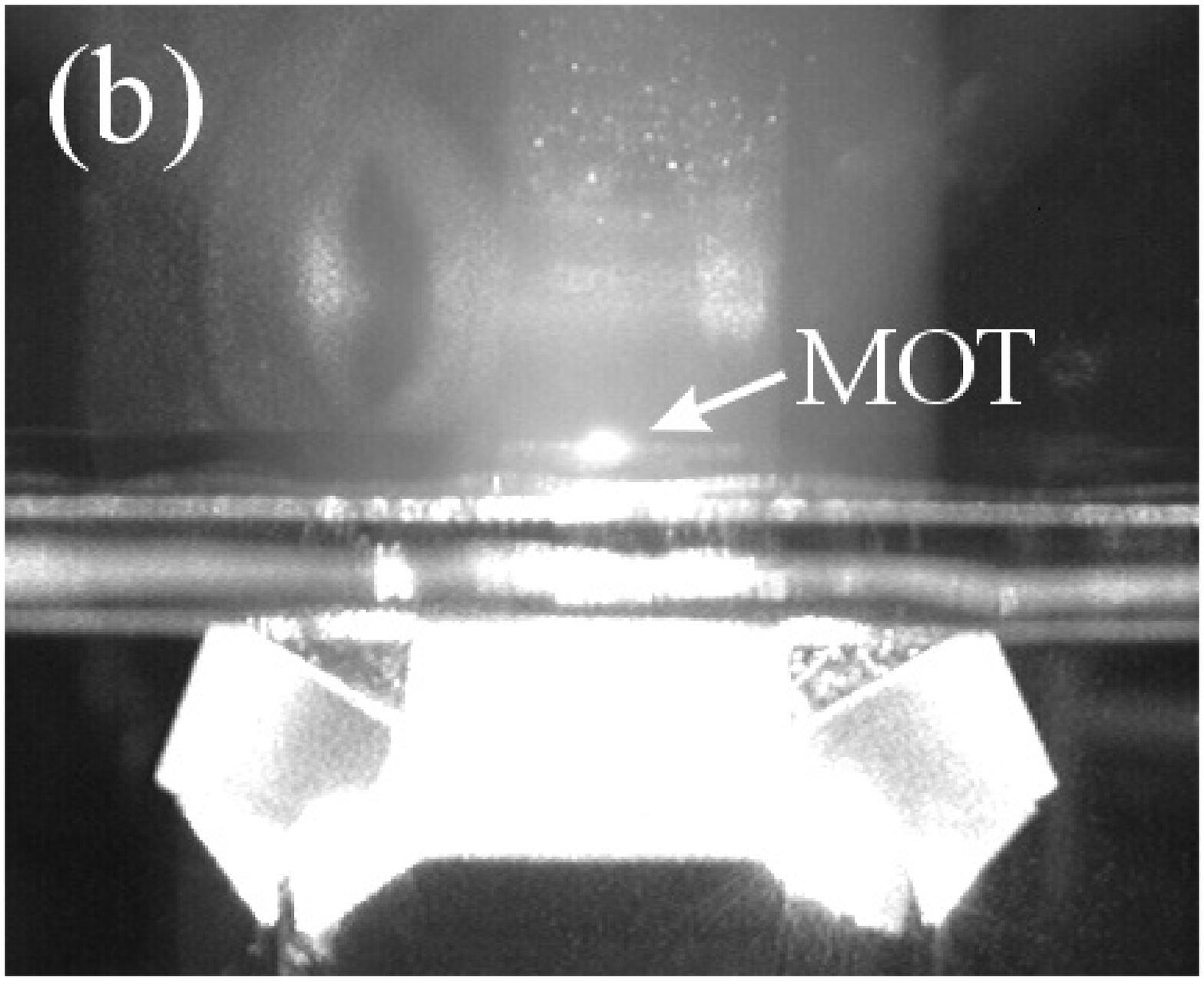}
\end{minipage}
\caption{\label{expt} A 3D schematic (a) of the demonstration setup using beams reflected from a single downward beam by three mirrors. The mirror
declination is $\theta=22.5^{\circ}$ resulting in upwards beam at an angle $45^\circ$ above the plane. The atoms (red cloud) form inside the vacuum
bordered by the quartz cell window. The experimental version is shown in (b), with the MOT indicated by a white arrow.}
\end{figure}

The mirrors of the triplet, situated below the vacuum cell, are $8\times 10$\,mm and steer a 1 inch diameter vertical beam into three upgoing beams, as
depicted in Fig.~\ref{expt}(a). For this demonstration setup, the triplet is positioned underneath the $4\,$mm thick quartz vacuum chamber. The angle of
the mirrors with respect to the horizontal plane were set to $22.5^\circ$, more acute than the ideal tetrahedral case, in order to form a MOT with a
larger supra-plane capture volume. Unlike the ideal case of in-vacuo mirrors, our resulting upwards beams are slightly elliptically polarised as the
intensities of the $p$ and $s$ polarisation drop by $9\%$ and $24\%$ respectively as a result of the four quartz surfaces experienced by the beam before
it returns to the MOT location.

A pinhole spatially filters the incident (downward) beam before it reaches the cell to achieve a good intensity balance and for improved uniformity in
reflected beam intensity. Appropriate up:down intensity balance is probably the main working condition of our one beam demonstration MOT, as too much
radiation pressure from the upwards beams completely counteracts the trapping nature of the downward beam. Since $\theta=22.5^\circ$ we require
sufficient attenuation $(I_{\rm up}/I_{\rm down}<57\%)$ for MOT operation, as shown by the black dash-dotted curve in Fig.~\ref{figth}.

The MOT requires a polarisation handedness shift upon reflection, and hence metal-coated  mirrors are ideal. By analysing the polarisation of reflected
circularly polarised beams as a function of incidence angle we found that many HR-coated $0-45^\circ$ incidence dielectric mirrors also satisfy this
criterion, and we used dielectric mirrors in our demonstration MOT. For microfabrication gold-coating is a better option, as the coating is less bulky
because it can be much thinner than the wavelength of light.

The tetrahedral configuration is used to cool $^{87}$Rb atoms via the D2 $F=2\rightarrow F'=3$ transition. The repumping light is generated by direct
modulation of the trapping laser diode current \cite{Myatt:1993:a}. As the energy splitting between the ground states is $6.835\,$GHz, we use a $13\,$dBm
output $6.5\,$GHz VTO-8650 Voltage Controlled Oscillator (VCO) from Avantek/PhaseMatrix to tune the modulation frequency and scan the two repumping
transitions $F=1\rightarrow F'=1$ and $F=1\rightarrow F'=2$. High-frequency boards, relaxation oscillations resonance \cite{Melentiev:2001:a} and optimum
external cavity size \cite{Myatt:1993:a} are used to maximize the repump light generation efficiency. For further details of frequency modulation in a
diode laser, see \cite{us}.

\begin{figure}[!ht]
\centering\includegraphics[width=8cm]{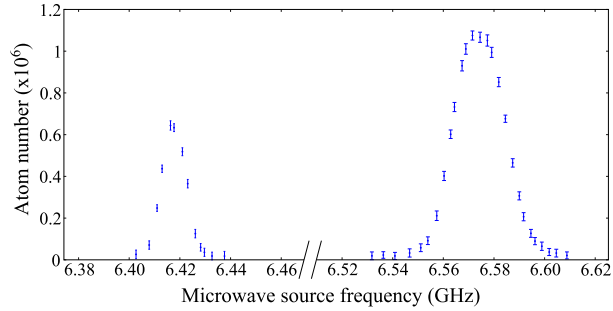} \caption{\label{figRF} Number of atoms with respect to the microwave frequency. The peaks correspond to
the $F=1\rightarrow F'=1$ (left) and $F=1\rightarrow F'=2$ (right) repumping transitions}
\end{figure}

Using the pyramid in conjunction with independent cooling and repumping lasers, fluorescence measurements indicated that $1.3\times 10^6$~atoms were
trapped. Fig.~\ref{figRF} shows the number of atoms in the MOT using the current modulation. The amount of repump light generated is of the order of
$0.5\,$mW, enough to get a maximum of $1.1\times
10^6$ atoms. 
The two peaks correspond to the repumping transitions $F=1\rightarrow F'=1$ and $F=1\rightarrow F'=2$. The positions of the maxima are in accordance with
the expected values for a detuning of the cooling light of 8.3\,MHz from the $F=2\rightarrow F'=3$ transition. We note also that using the
$F=1\rightarrow F'=2$ transition to repump the atoms is more efficient than the $F=1\rightarrow F'=1$ transition. This is to be expected as the branching
ratio to the $F=2$ level is more favourable via decays from $F'=2$ rather than the $F'=1$ level. Finally, it is worth noticing the broadening of the
repump transition, particularly visible on the rightmost peak ($\sim 25\,$MHz FWHM instead of $\sim 6\,$MHz).

The relatively small number of atoms in the MOT is due mainly to the small capture volume $(\approx 400\,$mm$^3$) of our demonstration setup. For
comparison, in a separate experiment in the same cell, an $8000\,$mm$^3$ 6-beam MOT collects $2\times 10^8$ atoms. As the atom number $N$ for a MOT with
capture volume $V$ scales approximately as $N\propto V^{1.2}$ \cite{lindquist}, we must only account for a factor of four in atom number reduction. This
reduction can be attributed to our experiment currently operating far from the optimal cooling/trapping regime of the pure tetrad configuration, and to a
lesser extent due to the reduced cooling and trapping of a 4-beam MOT compared to a 6-beam MOT. For smaller traps, uniform intensity beams and small
velocity/position changes imply a deceleration linear in velocity, $a=v dv/dx=-\alpha v$, leading to a capture velocity ${v_c}=\alpha d$ for beam
diameter $d.$ The atom number scales as $N\propto d^2 {v_c}^{4}$ \cite{lindquist} and thus $N\propto d^6,$ i.e $N=k_V V^{2}$ \cite{edprep}. As our design
obviates some of the constraints placed on the very first integrated atom-chip pyramid MOT \cite{edprep}, we predict the experimental proportionality
constant $k_V$ can be increased. For smaller MOTs optimal detuning decreases ($\Delta_\Gamma\approx -0.75$ for our parameters), and one can estimate the
rapid scaling with trap volume begins at pyramid sizes around $v/\alpha=0.6\,$mm since the damping is then linear for speeds $|v|<3\,$m/s and
$\alpha\approx 5000\,$s$^{-1}$). To circumvent the low background vapor loading rate for small pyramids, the use of loading mechanisms with low 3D
spatial and velocity spreads and zero final center-of-mass velocities could be used: e.g.\ magnetic transport \cite{magtrans} or magnetic lensing/optical
guiding \cite{lensing} of launched atoms.

\section{Outlook}
\label{speculate}

One immediate advantage of our pyramidal MOT is that, with balanced optical molasses, it should be possible to use a large (few cm) diameter pyramid as
the initial stage of an in-pyramid BEC experiment (using either magnetic trapping or supra-plane dipole beam trapping). The pyramid mirrors could
subsequently be used, in conjunction with a single focused dipole beam, to generate a phase- and geometry- stable 3D optical lattice for a `simple' Mott
insulator \cite{mott} experiment. A dipole beam waist diameter of several wavelengths (typical for ultracold atom experiments) would limit the effects of
edge-diffraction. The reflective single beam MOT/optical lattice can also be obtained using reflectors in a planar geometry -- i.e.\ three (or more)
separate gold-coated in-plane gratings blazed to give maximum output at the appropriate angle such that the intensity-weighted $\textbf{k}$ vectors yield
a balanced optical molasses. The planar grating geometry yields a MOT with no sub-plane capture volume (Fig.~\ref{volfig}(a)), however it greatly reduces
the volume of material etched/removed during nanofabrication, markedly reducing manufacture costs. Gratings with relatively low period, and
correspondingly low polarisation sensitivity, must be used to ensure the polarisation of the downwards beam simply flips handedness ($s$ reverses) after
diffraction.

Another exciting prospect is the use of the pyramid trap in conjunction with AFM tips, or other nano-indenters. These diamond nanodevices are extremely
hard and have tips sharp on a scale of tens of nanometres. Moreover, they have an adjustable pyramid angle, and could effectively be used as a `stamp' to
impress pyramidal shapes, potentially with smoothness at the atomic level, into almost any material. These impressions can be subsequently gold-coated
for reflection. With sufficient optical access to the base of a nano-indenter pyramid, an AFM tip itself could be used to form a refractive MOT, or
refractive optical lattice (Fig.~\ref{fig1} (b)) with precise control of relative cold atom-tip location. One can envisage mechanisms whereby one can
study the interaction between cold atoms and an AFM tip, reminiscent of the experimental demonstration of electron microscopy of BECs \cite{ott}. It is
even possible \cite{nanoindent} to use monolithic pyramid and cantilever devices out of (transparent) diamond, as cooled nanomechanical cantilevers and
their interaction with cold atoms \cite{reichel} present important new systems for studying, e.g. Casimir forces and non-Newtonian gravity
\cite{casnonnewt}.

\section{Conclusion}

We have demonstrated a pyramidal MOT using a single circularly-polarised beam and only three mirrors. No extra polarization optics are required. This
system has optimum performance when the beams cross in a perfect tetrahedral configuration. For our experimental requirements, the angle of the reflected
beams was increased in order to allow the formation of a novel supra-plane pyramid MOT. The design will be ideal for application as microfabricated
in-vacuo tetrahedral MOTs, a robust atomic beam source, a `simple' BEC/Mott insulator device, and a possible probe for non-Newtonian gravity.

\section{Acknowledgments}
We acknowledge valuable discussions with Ifan Hughes, Ian McGregor and Carol Trager-Cowan. Many thanks to Nicholas Houston for performing reflected-beam
polarization measurements.
\end{document}